\documentclass[pra, 12pt, a4paper, floatfix,superscriptaddress]{revtex4-2}

\usepackage{chemformula} 
\usepackage[T1]{fontenc} 
\usepackage{amssymb}
\usepackage{amsmath}

\usepackage{hyperref}
\usepackage{soul}

\begin{document}

\author{Ching-Hung Shih}
\affiliation
{Institute of Electronics, National Yang Ming Chiao Tung University, Hsinchu 300, Taiwan}

\author{Guan-Hao Peng}
\affiliation
{Department of Electrophysics, National Yang Ming Chiao Tung University, Hsinchu 300, Taiwan}

\author{Ping-Yuan Lo}
\affiliation
{Department of Electrophysics, National Yang Ming Chiao Tung University, Hsinchu 300, Taiwan}

\author{Wei-Hua Li}
\affiliation
{Department of Electrophysics, National Yang Ming Chiao Tung University, Hsinchu 300, Taiwan}

\author{Mei-Ling Xu}
\affiliation
{Department of Electrophysics, National Yang Ming Chiao Tung University, Hsinchu 300, Taiwan}

\author{Chao-Hsin Chien}
\email{chchien@nycu.edu.tw}
\affiliation{Institute of Electronics, National Yang Ming Chiao Tung University, Hsinchu 300, Taiwan}

\author{Shun-Jen Cheng}
\email{sjcheng@nycu.edu.tw}
\affiliation{Department of Electrophysics, National Yang Ming Chiao Tung University, Hsinchu 300, Taiwan}

\title {
Signatures of valley drift in the diversified band dispersions of bright, gray, and dark excitons in \ch{MoS2} monolayers under uni-axial strains
  }



\begin{abstract}

We present a comprehensive theoretical investigation of the excitonic properties of uni-axially strained transition-metal dichalcogenide monolayers (TMD-MLs) by solving the Bethe-Salpeter equation (BSE) in the Wannier tight-binding (WTB) scheme established on the base of first-principles. 
The presence of an uni-axial strain in a \ch{MoS2}-ML is shown to impact most the spin-forbidden exciton states, i.e. gray exciton (GX) and dark exciton (DX) states. 
As a consequence of strain-induced valley drift (VD) and electron-hole exchange interaction (EHEI), imposing an uni-axial strain onto a \ch{MoS2}-ML dramatically reshapes the band dispersion of DX from a parabola to a Mexican-hat-like profile, accompanying with unusual sign-reversal of the exciton effective mass and slight strain-activated brightness. 
Contrarily, the band dispersion of spin-forbidden GX yet remains parabolic but is characterized by a drastically reduced effective mass by an imposed strain.
The first-principles-based studies further predict the strain-induced diversification of the exciton diffusivities, transition dipoles, and angle-resolved optical patterns of BX, GX, and DX in uni-axially strained TMD-MLs, suggesting the feasibility of {\it spatially} resolving spin-allowed and -forbidden excitons in exciton  transport and optical spectroscopy experiments. 
\end{abstract}

\maketitle

\section{Introduction}

Transition metal dichalcogenide (TMD) monolayers (MLs) have been known as a promising low-dimensional material for valley-based photonic and excitonic applications.\cite{Tony-Heniz-2010,Yao-Wang-2015,Valley-polarization-1,valley-2,valley-3,Louie-2014,Shi-Su-Fei-2019, exciton-transport-device-1,exciton-transport-device-2,exciton-transport-device-3,exciton-device-1-twist-light,exciton-device-2-solar-cell,exciton-device-3-LED,exciton-device-4-QY} Because of enhanced electron-hole ({\it e-h}) Coulomb interactions in the 2D systems, photo-excited TMD-MLs exhibit pronounced exciton fine structures\cite{band-nesting,MacDonald-2015,Pawel-2022}, allowing for spectrally resolving exciton complexes in variety, including spin-allowed bright excitons (BXs) and spin-forbidden gray (GX) and dark excitons (DXs). {\color{black}\cite{X-Marie-2016,WH-2023,shao-yu-2024,Erfu-Liu-PRL,Chun-Hung-liu-PRL}} 
In particular, atomically thin TMD-MLs present high mechanical flexibility that can withstand the strain so high as 10$\%$,\cite{strain-limit}   suggesting the prospect of TMD-based straintronics.\cite{straintronic}
To date, effective mechanical controls of the electronic and optical properties of TMD-MLs, comprising \ch{WSe2}, \ch{MoS2}, \ch{WSe2} and \ch{MoSe2} monolayers, have been demonstrated by applying bi-axial \cite{1-biaxial,2-biaxial,3-biaxial}, uni-axial \cite{2013-Nano-lett,1-uniaxial,2-uniaxial}, and non-uniform strains as well.\cite{1-nonuniform,2-nonuniform}
Remarkably, imposing a uni-axial strain to a crystalline material can fundamentally transform the inherent crystal symmetry and often gives rise to peculiar electronic and excitonic structures, such as
direct-indirect band gap transitions, \cite{2013-Nano-lett, 2-raman,3-raman} strain-induced fine structure splitting of BX \cite{experiment-splitting,experiment-splitting-2}, strain-guided exciton transport,\cite{Malic-2022,exciton-diffusion-1} formation of quasi-1D localized exciton, \cite{Quasi-1D} and valley drift (VD) \cite{valley-orbital-magnetization-2019,valley-orbital-magnetization-2017}.

Regarding the strain-induced VD, the band edge states of conduction and valence bands of a uni-axially strained TMD-ML was predicted by the density functional theory (DFT) to be relocated apart from the K- and K$^\prime$-points in the reciprocal space.\cite{valley-drift-2013,valley-drift-2019,valley-drift-2021}
Such a strain-induced VD has recently drawn considerable attention because of its essential role in strain-induced Berry curvature dipoles, leading to non-linear Hall conductivities \cite{Berry-cuvature-2015} and valley orbital magnetization as experimentally observed in {\color{black}Refs.~\cite{Berry-cuvature-2018, valley-orbital-magnetization-2019,valley-orbital-magnetization-2017}.}
However, the effects of strain-induced VD on the optical and excitonic properties of TMD-MLs remain rarely explored so far.

Theoretically, the first theoretical research on the BX band structures of uni-axially strained \ch{MoS2}-MLs could be traced back to the pioneering work by Hongyi Yu {\it et al.} \cite{Yao-Wang-2014,Yao-Wang-2015}, which predicted the strain-induced splitting of BX doublet at zero momentum and {\color{black}crossings} of longitudinal and transverse BX bands at finite momenta. Recently, M. M. Glazov {\it et al.} built up a strain-dependent model Hamiltonian for BXs in a \ch{WSe2}-ML under uni-axial strain according to the symmetry arguments. \cite{PRB-2022-splitting, main-PRB-2022-splitting} 
Yet, the excitonic effects of VD  in uni-axially strained TMD-MLs, especially on spin-forbidden GX and DX, have not been investigated and identified on the base of first principles. 

In this work, we present a comprehensive theoretical investigation of the strain-modulated excitonic properties of uni-axially strained \ch{MoS2}-MLs by solving the Bethe-Salpeter equation (BSE) in the Wannier tight-binding {\color{black} (WTB)} scheme established on the basis of first principles.
{\color{black} In the WTB scheme, we develop an efficient computation methodology for solving the first-principles-based BSE in the basis of maximally localized Wannier function (MLWF). By taking advantages of analytic integration based on spline interpolation, the electron-hole Coulomb matrix elements of the BSE are efficiently and accurately calculated.   }
We show that imposing an uni-axial strain onto a \ch{MoS2}-ML results in diversified fine structures and band dispersions of BX, GX, and DX states, as a consequence of the competitive interplay between strain-induced VD and momentum-dependent {\it e-h} exchange interaction (EHEI). 
The first-principles-based studies suggest that the strain-diversified band dispersions, characterized by very distinct effective exciton masses, of BX, GX and DX  makes it possible to not only {\it spectrally} but also {\it spatially} resolve BX, GX and DX in the exciton diffusive transport experiments or angle resolved optical spectroscopy on uni-axially strained TMD-MLs.

\section{Theory}
\subsection{{\color{black} Electronic band structures in the density functional theory (DFT)} }
{\color{black} First, we calculate the quasi-particle band structure of a \ch{MoS2}-ML by solving} the Kohn-Sham (KS) equation in the DFT, 
 \begin{equation}
H^{\rm{KS}} \lvert \psi_{n,\boldsymbol{k}} \rangle = \epsilon_{n,\boldsymbol{k}} \lvert \psi_{n,\boldsymbol{k}} \rangle ,
\end{equation}
where $\epsilon_{n,\boldsymbol{k}}$ $ (\psi_{n,\boldsymbol{k}})$ is the eigen-energy (wave function) of the solved Bloch states of the $n$-th band with the wave vector $\boldsymbol{k}$. 
{\color{black} The DFT calculations are performed by using the package of Quantum Espresso (QE) code \cite{QE}, with the adoption of the Perdew-Burke-Ernzerhof (PBE)\cite{PBE} model for the exchange-correlation functional and the consideration of the spin-orbit interaction. }
{\color{black}
Throughout this work, we consider \ch{MoS2}-MLs with uni-axial stresses along the $x$-axis, and adopt the strain tensor imposed to the lattices of un-strained \ch{MoS2}-ML in the DFT simulations,
\begin{equation}
\hat{\varepsilon} =
\begin{bmatrix}
    \varepsilon_{xx} &  0 & 0\\ 
    0 & \varepsilon_{yy} & 0\\ 
    0 &            0                  &0
\end{bmatrix},
\end{equation}
with the neglect of the small $\varepsilon_{zz}$ for 2D materials. Taking the Poisson's ratio $\nu=0.31$ for a \ch{MoS2}-ML, we have $\varepsilon_{yy}= -\nu\varepsilon_{xx}=-0.31 \varepsilon_{xx} $ \cite{Ctensor-PRB-2013}.
}
 More technical details of the DFT calculations for strained \ch{MoS2}-MLs are described by {\color{black}Appendix A.} 
\
\subsection{{\color{black} DFT-WTB-BSE theory of exciton } }
\label{BSE-theory}

On the basis of the DFT-calculated electronic band structures of a 2D material, one can express the wave function of an exciton state with  the exciton wave vector, $\boldsymbol{Q}=(Q_x,Q_y)$, as
\begin{equation}
|S , \boldsymbol{Q} \rangle=\frac{1}{\sqrt{\mathcal{A}}} \sum_{vc\boldsymbol{k}}A_{S,\boldsymbol{Q}}\left(vc\boldsymbol{k}\right)\hat{c}^{\dag}_{c,\boldsymbol{k}+\boldsymbol{Q}}\hat{h}^{\dag}_{v,-\boldsymbol{k}}|GS \rangle, 
\end{equation}
being a linear combination of electron-hole-pair configurations, $\hat{c}^{\dag}_{c,\boldsymbol{k}+\boldsymbol{Q}}\hat{h}^{\dag}_{v,-\boldsymbol{k}}|GS \rangle$,
where $S$ is the exciton band index, $v$($c$) is the valance (conduction) band index, $\hat{c}^{\dag}_{c,\boldsymbol{k}}(\hat{h}^{\dag}_{v,-\boldsymbol{k}})$ is the creation operator for an electron (a hole), $|GS \rangle$ denotes the many-body ground state, the coefficient $A_{S,\boldsymbol{Q}}\left(vc\boldsymbol{k}\right)$ as a function of $\boldsymbol{k}$ can be treated as the wave function of exciton in the reciprocal space, and $\mathcal{A}$ is the area of the \ch{MoS2}-ML.
$A_{S,\boldsymbol{Q}}\left(vc\boldsymbol{k}\right)$ is obtained by solving the BSE, which reads
\begin{equation}
\left[\epsilon_{c,\boldsymbol{k}+\boldsymbol{Q}}-\epsilon_{v,\boldsymbol{k}}\right]A_{S,\boldsymbol{Q}}\left(vc\boldsymbol{k}\right)+\sum_{v^{\prime}c^{\prime}\boldsymbol{k}^{\prime}}U_{\boldsymbol{Q}} \left(vc\boldsymbol{k},v^{\prime}c^{\prime}\boldsymbol{k}^{\prime} \right)A_{S,\boldsymbol{Q}}\left(v^{\prime}c^{\prime}\boldsymbol{k}^{\prime}\right)=E^{X}_{S,\boldsymbol{Q}}A_{S,\boldsymbol{Q}}\left(vc\boldsymbol{k}\right),
\label{BSE}
\end{equation}
where the Coulomb kernel, $U_{\boldsymbol{Q}}\left(vc\boldsymbol{k},v^{\prime}c^{\prime}\boldsymbol{k}^{\prime} \right) =-V^d_{\boldsymbol{Q}}\left(vc\boldsymbol{k},v^{\prime}c^{\prime}\boldsymbol{k}^{\prime} \right)+V^x_{\boldsymbol{Q}}\left(vc\boldsymbol{k},v^{\prime}c^{\prime}\boldsymbol{k}^{\prime} \right) $ , is composed of the electron-hole direct (former term) and exchange (latter one) interactions. The matrix element of direct Coulomb interaction is defined by
\begin{equation}
V_{Q}^{d}(vc\boldsymbol{k}, v'c'\boldsymbol{k}') = \int d^{3}\boldsymbol{r}_1 d^{3}\boldsymbol{r}_2 \, \psi_{c,\boldsymbol{k}+\boldsymbol{Q}}^{*}(\mathbf{r}_1) \psi_{v,\boldsymbol{k}}(\mathbf{r}_2) {\color{black}W(\boldsymbol{r}_1, \boldsymbol{r}_2)} \psi_{v',\boldsymbol{k}'}^{*}(\boldsymbol{r}_2) \psi_{c',\boldsymbol{k}'+\boldsymbol{Q}}(\boldsymbol{r}_1) , 
\label{direct_Coulomb_interaction}
\end{equation}
and that of electron-hole exchange interaction (EHEI) is
\begin{equation}
V_{Q}^{x}(vc\boldsymbol{k}, v'c'\boldsymbol{k}') = \int d^{3}\boldsymbol{r}_1 d^{3}\boldsymbol{r}_2 \, \psi_{c,\boldsymbol{k}+\boldsymbol{Q}}^{*}(\boldsymbol{r}_1) \psi_{v,\boldsymbol{k}}(\boldsymbol{r}_1) V(\boldsymbol{r}_1-\boldsymbol{r}_2) \psi_{v',\boldsymbol{k}'}^{*}(\boldsymbol{r}_2) \psi_{c',\boldsymbol{k}'+\boldsymbol{Q}}(\boldsymbol{r}_2) , 
\label{EHEI}
\end{equation}
where $V(\boldsymbol{r}_1, \boldsymbol{r}_2) = \frac{e^2}{4\pi\epsilon_0 |\boldsymbol{r}_1 -\boldsymbol{r}_2|}$ is the bare Coulomb interaction, ${\color{black}W(\boldsymbol{r}_1,\boldsymbol{r}_2)} = \int d^3\boldsymbol{r}' \, {\color{black}\epsilon^{-1}(\boldsymbol{r}_1, \boldsymbol{r}')} V(\boldsymbol{r}' - \boldsymbol{r}_2)$ is the screened Coulomb interaction, 
and ${\color{black}\epsilon^{-1}(\boldsymbol{r}_1, \boldsymbol{r}_2)}$ is the inverse {\color{black}q-dependent} dielectric function. 
{\color{black}Following previous literature \cite{sci-2017,Diana-2016,Thygesen-2015}, we neglect the local field effect  \cite{local-field}. Thus, we adopt the approximation $W(\boldsymbol{r}_1,\boldsymbol{r}_2)\approx W(\boldsymbol{r}_1-\boldsymbol{r}_2)$ and $\epsilon^{-1}(\boldsymbol{r}_1,\boldsymbol{r}_2)\approx \epsilon^{-1}(\boldsymbol{r}_1-\boldsymbol{r}_2) $.}

{\color{black} 
{\color{black} Since the matrix elements of Coulomb kernel essentially involve the rapidly varying microscopic parts of the Bloch wave functions,
the precise evaluation of the Coulomb kernel, especially $V^x_{\boldsymbol{Q}}\left(vc\boldsymbol{k},v^{\prime}c^{\prime}\boldsymbol{k}^{\prime} \right) $, is computationally non-trivial. 
To overcome the numerical difficulties, we alternatively re-formulate the DFT and BSE in the WTB scheme, which has been shown advantageous in the reduction of numerical costs.\cite{GH-2019}
The Wannier functions are obtained by means of the transformation from the DFT-calculated Bloch states, which is defined by
{\color{black}
\begin{equation}
\vert \mathcal{W}_{j,\mathbf{R}} \rangle = \frac{1}{\sqrt{N}}\sum_{\boldsymbol{k}}e^{-i\boldsymbol{k}\cdot \boldsymbol{R}}\sum_{n=1}^{N_b}\big[U^{(\boldsymbol{k})}\big]_{nj}| \psi_{n,\boldsymbol{k}}\rangle,
\end{equation}
where $\boldsymbol{R}$ is the Bravais lattice vector, $j$ = ($J$,$\alpha$) is the composite index that denotes the atomic orbitals ($\alpha=p_x,p_y,p_z,d_{x^2-y^2},...$) of the $J$-th atom in the primitive cell, $N$ is the total number of primitive cells in the system, $N_{b}$ is the total number of electornic bands, and $U^{(\boldsymbol{k})}$ is the matrix for the unitary transform of Wannier function that is implemented by using the WANNIER90 package.
}{\color{black}\cite{wannier90}}

Taking the MLWFs as basis, one can rewrite the KS Hamiltonian as $H^{\rm{TB}}_{ij}(\boldsymbol{k})= \sum_{\mathbf{R}} e^{i \boldsymbol{k} \cdot \mathbf{R}} t_{ij}^{w}(\mathbf{R})$ in the form of WTB-Hamiltonian matrix, where $t_{ij}^{w}(\mathbf{R}) \equiv \langle \mathcal{W}_{i,\boldsymbol{0}} | H^{KS} | \mathcal{W}_{j,\mathbf{R}} \rangle$. 
The DFT-calculated Bloch wave function in the WTB scheme is re-written as
\begin{equation}
|{\psi_{n,\boldsymbol{k}}}\rangle = \sum_{j=1}^{N_b} C_j^{\left(n\right)} \left(\boldsymbol{k}\right) |{\phi_{j,\boldsymbol{k}} }\rangle =\frac{1}{\sqrt{N}} \sum_{j=1}^{N_b}\sum_{\boldsymbol{R}} C_j^{\left(n\right)} \left(\boldsymbol{k}\right) e^{i\boldsymbol{k}\cdot \boldsymbol{R}}  |{\mathcal{W}_{j,\boldsymbol{R}} }\rangle,
\label{phi_bloch}
\end{equation}
where $ | {\phi_{j,\boldsymbol{k}} }\rangle=\frac{1}{\sqrt{N}}\sum_{\boldsymbol{R}}e^{i\boldsymbol{k}\cdot \boldsymbol{R}}|{\mathcal{W}_{j,\boldsymbol{R}}}\rangle$ is referred to as the Bloch sum function, and $\left\{ C_j^{\left(n\right)} \left(\boldsymbol{k}\right) \right\}$ is the eigenvector of $H^{\rm{TB}}_{ij}(\boldsymbol{k})$ to be determined by directly diagonalizing the WTB-Hamiltonian matrix.
 {\color{black} Substituting Eq.~\ref{phi_bloch} back into Eqs.~\ref{direct_Coulomb_interaction}  and ~\ref{EHEI}, we rewrite $V^d_{\boldsymbol{Q}}$ and $V^x_{\boldsymbol{Q}}$ in terms of Wannier functions, which are suitable for numerical calculations and are presented in Eqs. S1 and S2 of the Supplemental Material (SM). For more details about numerical estimation of the direct Coulomb interaction and EHEI in the WTB scheme, one can refer to SM.\cite{SM}}
}

{\color{black} In the WTB scheme, the DFT-based BSE can be established and solved in a efficient manner, which benefit the simplicity of the matrix form of WTB-Hamiltonian and the efficient evaluation of the Coulomb kernel by using the integration algorithm as presented in Ref.~\citenum{GH-2019}.
The high efficiency of our developed WTB-BSE solver enables the efficient calculations of the delicate strain-modulated exciton band dispersions of strained TMD-MLs over extended exciton momentum space.  

From the solved quasi-particle band structures, Bloch wave functions, and wave functions of exciton, the transition dipoles, $\boldsymbol{D}^X_{S,\boldsymbol{Q}}$, of an exciton in the $\vert S,\boldsymbol{Q} \rangle$ states of strained \ch{MoS2}-MLs that play a vital role in the light-matter interaction can be calculated using the formalisms presented in {\color{black}Appendix B}.
}

Following the approach of Refs.~\citenum{WH-2023, sci-2017}, we solve the {\color{black}q-dependent} dielectric function of a TMD-ML sandwiched by h-BN layers using the semi-classical electromagnetic theory. 
{\color{black} In the approach, we consider an electron-hole pair in a TMD-ML embedded in a hBN bulk (forming a hBN/TMD/hBN 3-layer system) and solve the Poisson's equation for the screened Coulomb potential experienced by the charged particles. The dielectric properties of the multi-layer dielectrics are characterized with piece-wise dielectric constants for hBN ($\varepsilon_{\text{hBN}}=5.89$) and the three-dimensional dielectric function for TMD materials developed by Refs.\citenum{1993-dielectric, sci-2017},
\begin{equation}
\varepsilon_{\text{TMD}} (q) = 1 + \left[ \frac{1}{\varepsilon_{\parallel}^{bulk} -1} +\alpha \frac{q^2}{q_{TF}^2} + \left( \frac{\hbar^2 q^2}{2m_0 E_{PL}} \right)^2   \right]^{-1}
\label{dielectric_func_TMD}
\end{equation}
where $m_0$ is free electron mass,  $\varepsilon_{\parallel}^{bulk}=13.2$ is the in-plane dielectric constant of \ch{MoS2} bulk,  $q_{TF}=\sqrt{ \frac{e^2 m_0}{\pi^2 \varepsilon_0 \hbar^2} \left( \frac{3 \pi^2 \varepsilon_0 m_0 E_{\text{pl}}^2}{e^2 \hbar^2} \right)^{1/3} }=23.1\text{nm}^{-1}$ is the Thomas-Fermi wave vector, $E_{PL}=22.5$eV is the plasma peak energy, and $\alpha=1.55$ is a fitting factor.\cite{sci-2017, 2020-dephasing} After solving the screened Coulomb interaction energy, the {\color{black}q-dependent} dielectric function is obtained from the ratio of the $z$-averaged screened Coulomb interaction to the bare one. \cite{WH-2023}  
From Ref.\citenum{WH-2023} (See Fig.3 therein), one realizes that the non-local effect in the q-dependent dielectric function affect the magnitude of the fine structure splitting between BX and DX bands because of their unequal Bohr radii. For a \ch{MoS2}-ML, neglecting the non-locality of dielectric screening and taking the dielectric function in the long-range approximation could overestimate the exciton fine splittings.

For studying excitons in strained TMD-MLs, one should in principle consider the strain effects on the dielectric function of Eq.\ref{dielectric_func_TMD}. 
Yet, the strain-dependences of {\color{black}q-dependent} dielectric functions for TMD-MLs remain a rarely explored subject in the literature.\cite{epsilon_2D}
Considering that only only the small wave vectors, $q\sim a_B^{-1} \ll q_{TF}=23.1\text{nm}^{-1}$, are relevant to a Wannier exciton with $a_B\sim 1- 2 \text{nm}$ in a TMD-ML, $\varepsilon_{\parallel}^{bulk}$ acts as the leading term in Eq.\ref{dielectric_func_TMD}. 
Experimentally, Ref.~\citenum{epsilon_2D} showed that $\varepsilon_{\parallel}^{bulk}$ is changed slightly by $1- 2\%$ as imposing a $1\%$ bi-axial strain to a \ch{MoS2} bulk ($\varepsilon_{\parallel}^{bulk}$ is increased from 16.03 to 18.98 with increasing the imposed biaxial strain from $0 \%$ to $10\%$). Thus, throughout this work we assume the negligible strain-dependence of dielectric function and adopt the fixed parameters in Eq.\ref{dielectric_func_TMD} for excitons in both strained and un-strained \ch{MoS2}-MLs . 
}



\section{Results and discussion}
\begin{figure} 
\centering 
\includegraphics[width=\textwidth]{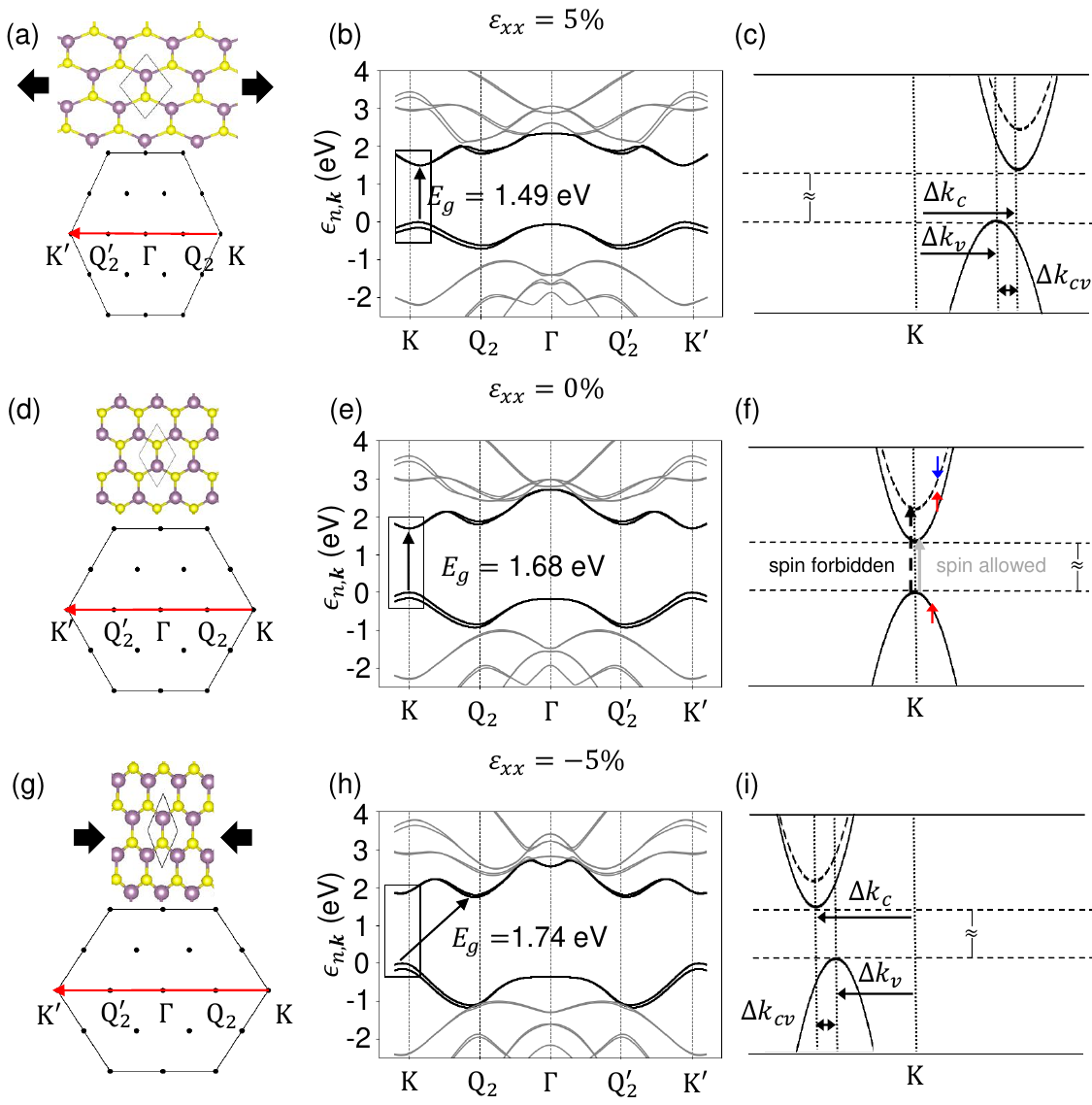}

\caption{ (a) Schematics of the lattice structures and reciprocal Brillouin zones of a \ch{MoS2}-ML with a tensile uni-axial strain ($\epsilon_{xx}>0$) along the x-axis, where the $x$-axis is specified to be along the axis of zigzag atom chain of \ch{MoS2}-ML.   (b) The DFT-calculated electronic band structures of a \ch{MoS2}-ML with the tensile uni-axial strain of $\varepsilon_{xx} = 5\%$. (c) The zoom-in view of the conduction and valence bands of (b), around the $\rm{K}$-valley. The solid (dashed) lines represent the up-spin (down-spin) bands. $\Delta k_c$ ($\Delta k_v$) denotes the strain-induced VD of the conduction (valence) band. (d-f) [(g-i)] Same as (a-d) but for the un-strained [compressively strained] \ch{MoS2}-ML with $\varepsilon_{xx} = 0\%$ [$\varepsilon_{xx} = -5\%$].}
\label{Fig. 1} 
\end{figure}

\subsection{Electronic band structure}

{\color{black}Fig.  1 (a), (d) and (g)} depict the lattice structures and the reciprocal Brillouin zones of a \ch{MoS2}-ML under a tensile uni-axial stress, no stress, and a compressive uni-axial stress, respectively. 
{\color{black} Experimentally, a common way to impose a controlled uni-axial strain, typically $1 \sim 2\%$,  in a TMD-MLs is by means of bending the flexible substrate (polyethylene terephthalate (PET)) that holds the TMD-ML.  Alternatively, larger uni-axial strains up to $5\%$ in atomically thin TMDs can be generated by using single-axis translation stage, as demonstrated by Ref.~\citenum{strain-0.05} in the home-made linear displacement set-up. }
Throughout this work, we consider the uni-axial stresses, $\pm 5\%$, applied to a \ch{MoS2}-ML along the crystalline axis of zigzag atom chain, specified to be the $x$-axis.
{\color{black}Fig.  1 (b), (e), and (h)} show the DFT-calculated quasi-particle band structures of a \ch{MoS2}-ML under the tensile uni-axial stress, no stress, and compressive uni-axial stress along the $x$-axis, yielding the uni-axial strains, $\varepsilon_{xx}=+5\%$, $0\%$, and $-5\%$, respectively. 


It is shown that, with varying the uni-axial strain from $\varepsilon_{xx}=5\%$, $\varepsilon_{xx}=0\%$ to $\varepsilon_{xx}=-5\%$, the energy gap of the strained \ch{MoS2}-ML increases from $E_g=1.49eV$, $E_g=1.68eV$, to $E_g=1.75eV$, and transits from an intra-valley ($\rm{K}$-$\rm{K}$) energy gap to an inter-valley ($\rm{K}$-$\rm{Q}$) indirect gap.
Without employing the GW method,\cite{BGW-1986,BGW-2012,GPAW-2010} the calculated band gaps in the DFT are underestimated as compared with experimental values.{\color{black} \cite{Tony-Heniz-2010,PL-2010-1}} Nevertheless, the predicted strain-dependent electronic and excitonic structures of \ch{MoS2}-MLs by the DFT with the broadly adopted {\color{black} PBE} functional are sufficiently valid for the physical investigation under the scope of this work.
{\color{black}Fig.  1 (c), (f) and (i)} shows the zoom-in views of the conduction and valence bands of the uni-axially strained \ch{MoS2}-ML around the $\rm{K}$ valley, showing the VD inwards (outwards) the center of the Brillouin zone under an imposed tensile (compressive) uni-axial strain. 
With $\varepsilon_{xx}=5\%$, the calculated drift of the conduction (valence) valley is given by $\Delta k_c= 0.091 \mathring{A}^{-1} = 94.8Q_c$ ($\Delta k_v= 0.074 \mathring{A}^{-1} = 76.8Q_c$), two orders of magnitude larger than the wave vector of light cone (LC) edge, $Q_c \equiv  \frac{E_{S,\bf{0}}^X}{\hbar c} \approx 9.6 \times 10^{-4}  \mathring{A}^{-1}$, where $\hbar$ denotes the Planck's constant, $c$ the speed of light in vacuum, and {\color{black} $E_{S,\bf{0}}^X \approx 1.9\rm{eV}$} is estimated by the DFT-BSE-calculated exciton energy.\cite{Diana-2013} Such uni-axial strain-induced VDs \cite{valley-drift-2013,valley-drift-2019,valley-drift-2021,valley-drift-2015} are recently recognized to be essential in non-linear topological physics and valley magnetization. \cite{topological-1,valley-orbital-magnetization-2019,valley-orbital-magnetization-2017,Berry-cuvature-2015,Berry-cuvature-2018,PRM-2023} 

Note that the strain-induced VDs of conduction and valence bands are towards the same direction but differ in the magnitude, i.e. $\Delta k_c \neq \Delta k_v$ for a fixed strain. The difference between the strain-induced conduction- and valence-VDs is approximately $3 Q_c/\%$, indicating a significant strain-induced momentum drift of exciton band.  Naively disregarding the Coulomb interaction of exciton, the lowest excited states of a uni-axially strained TMD-ML therefore would be the momentum-forbidden DX states with a large exciton wave vector, $Q= \Delta k_c - \Delta k_v \equiv \Delta k_{cv}$, out of the LC. However, as we shall show later, the competitive interplay between the VD and momentum-dependent EHEI leads to the diversified band dispersions of BX, GX and DX of an uni-axially strained TMD-ML, generally deviating from the naive non-interacting scenario.

\begin{figure}
\centering 
\includegraphics[width=\textwidth]{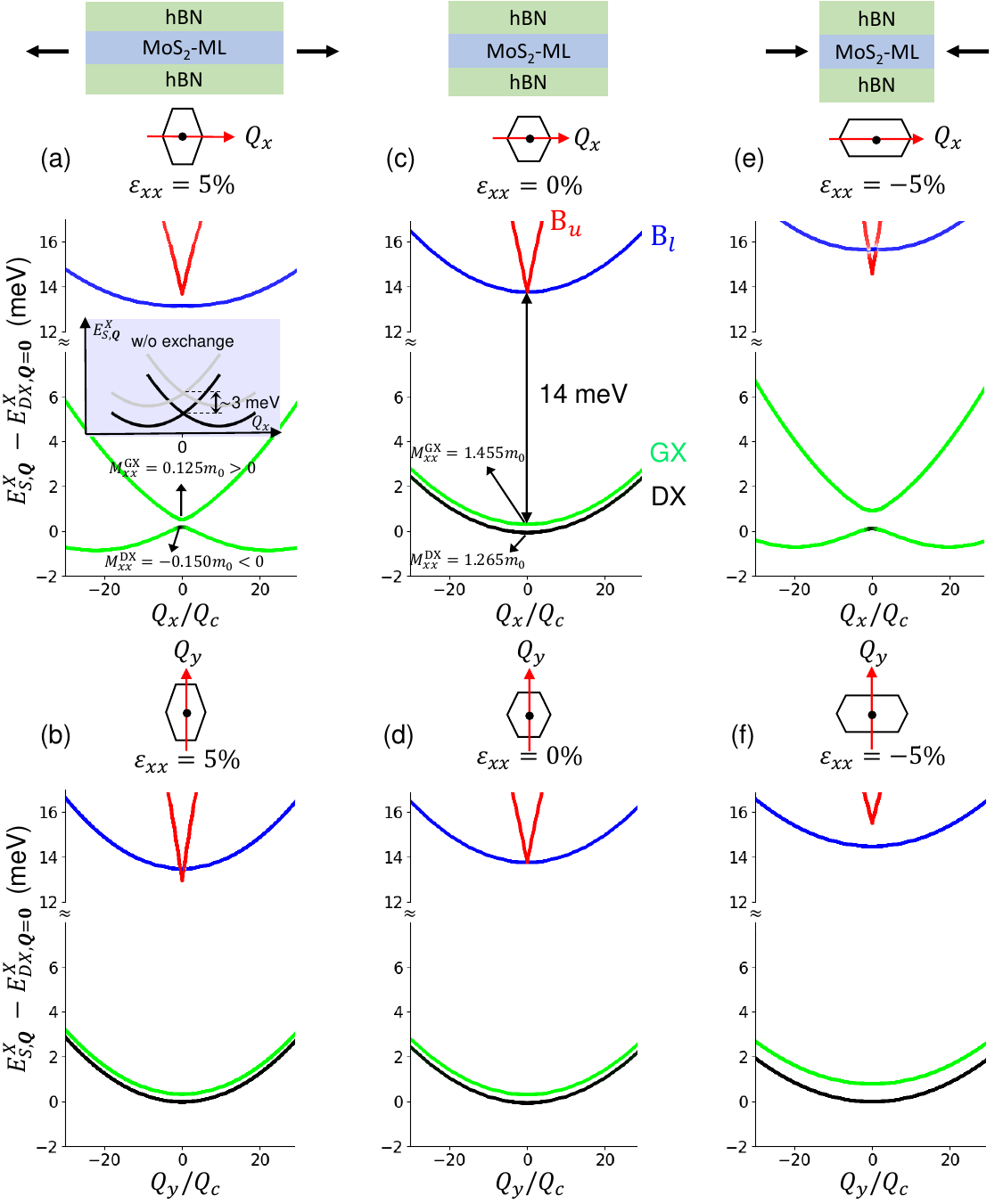} 
\caption{{\color{black}(a-b) Fine structure in the energy bands of BX, GX and DX in a strained \ch{MoS2-ML} with the tensile uni-axial strain, $\varepsilon_{xx} = +5\%$, along (a) $Q_{x}$-axis (b)$Q_{y}$-axis. The BX doublet is split by EHEI into the upper and lower bands, denoted by B$_u$ and B$_l$, respectively. The grading of red (blue) color represents the magnitude of longitudinal (transverse) transition dipoles. The grading of the color from dark to green represents the magnitude of the out-of-plane dipole of exciton.   Top inset: Schematic of the hBN-sandwiched \ch{MoS2}-MLs. (c-d) Same as (a-b) but for the un-strained \ch{MoS2}-ML. (e-f) Same as (a-b) but for the compressive uni-axial strain \ch{MoS2}-ML.} {\color{black}Inset in (a): Calculated exciton bands, with no consideration of EHEI, of the spin-allowed (gray) and spin-forbidden (black) exciton states of the \ch{MoS2-ML} under the tensile uni-axial strain.}}
\label{Fig. 2}
\label{Fig. 2b}  
\end{figure}

\begin{figure}
\centering
\includegraphics[width=1\textwidth]{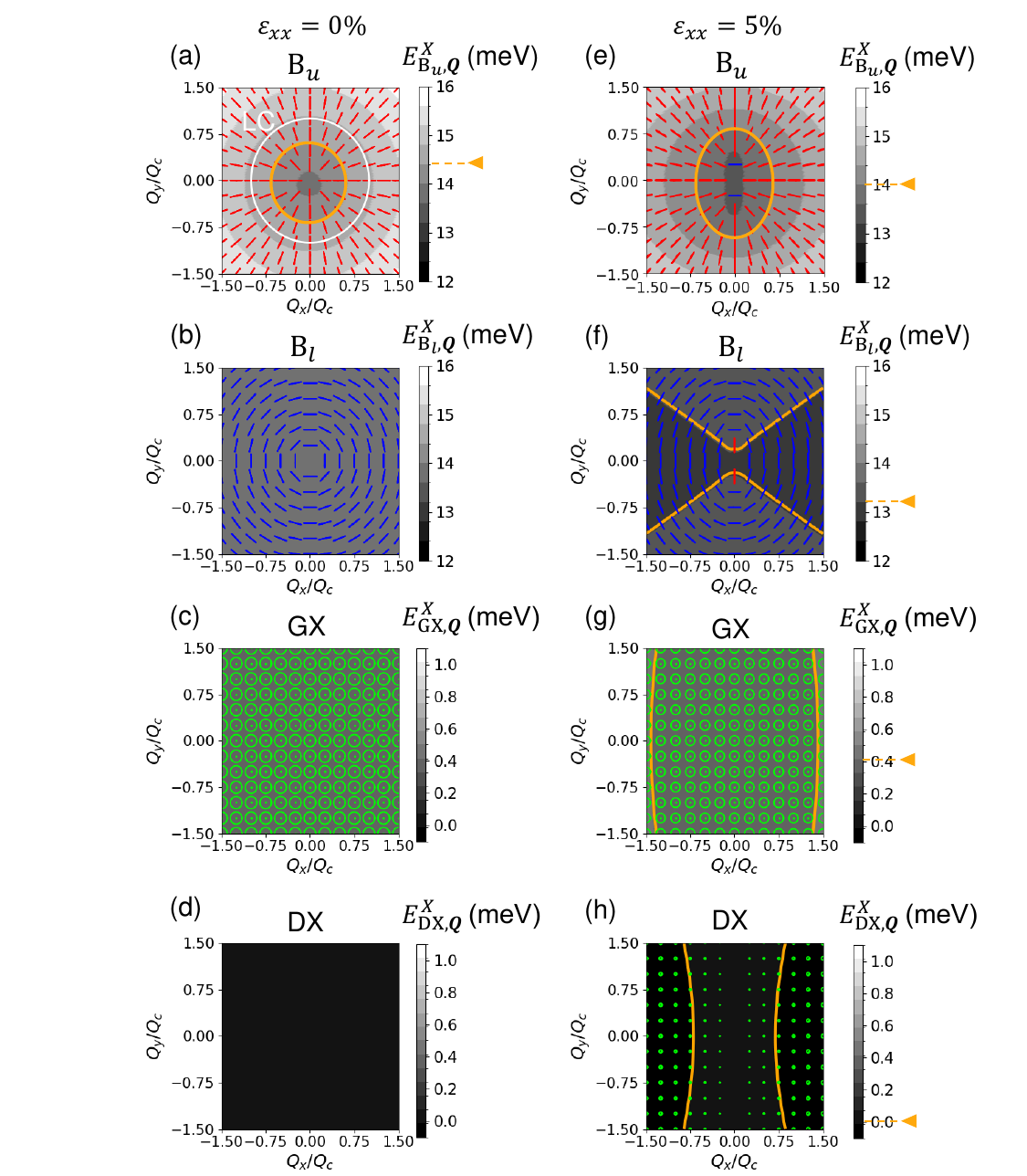}

\caption{ (a-d): The energy contours of the B$_u$, B$_l$, GX, and DX bands of an un-strained MoS$_2$-ML with $\varepsilon_{xx} = 0\%$. The red (blue) line segments placed in the $\boldsymbol{Q}$-plane represent the magnitudes and orientations of the longitudinal (transverse) dipoles of the BX states with $\boldsymbol{Q}$. The size of green circle represents the magnitude of out-of-plane dipole of GX and DX states.  (e-h): Same as (a-d) but for a uni-axially
strained \ch{MoS2}-ML with $\varepsilon_{xx} = 5\%$. The large white circle indicates the boundary of LC where $Q=Q_c$.  {\color{black}The orange-coloured energy contours illustrate the strain-induced anisotropies in the exciton band dispersion of MoS2-ML.}
} 
\label{Fig.  3}
\label{Fig.  3b} 
\end{figure}
\par

\subsection{Exciton band dispersions}

Based on the DFT-calculated electronic band structures, we proceed with the calculations of the fine structures and band dispersions of $1s$ exciton of the un-strained and strained {\color{black} hBN-encapsulated} \ch{MoS2}-MLs by using the  computational methodology of Sec. \ref{BSE-theory}.\cite{GH-2019,WH-2023}. 
For a 2D material system, the states of exciton are characterized by the in-plane wave vector in the center-of-mass coordinate, ${\boldsymbol{Q}}=(Q_x,Q_y)$, and labelled by the symbol $S$=($\text{B}_u,\text{B}_l, \text{GX}, \text{DX}$) to denote the type of exciton  {\color{black} according to the relative orientations of electron and hole spin}. {\color{black}   The exciton states in which the spins of electron and hole are {\it anti-parallel} to each other (i.e. {\it parallel} spins of the conduction electron and the missing valence electron) are referred to as bright exciton (BX) states that possess in-plane transition dipoles.  Throughout this work, $\text{B}_u$ and $\text{B}_l$ denote the BX states in the upper and lower bands that are split by the long-ranged EHEI. Contrarily, the exciton states with parallel electron and hole spins (i.e. {\it anti-parallel} spins of the conduction electron and the missing valence electron) are referred to as spin-forbidden dark exciton states, which are supposedly optically inactive and forbidden for optical transitions. However, due to the spin-orbit interaction in TMD-monolayers lacking the spatial inversion symmetry, one of the spin-forbidden exciton doublet is spin-mixed so as to possess weak out-of-plane dipoles and, by convention, named as gray exciton (GX).\cite{1-D-G-mixing} Yet, the other spin-forbidden exciton still remains totally dark and named as dark exciton (DX). With a weak out-of-plane dipole, GX is subjected to a weak repulsive short-ranged EHEI and normally is higher in energy than DX.}  
  
The eigenenergy, $E_{S,{\boldsymbol{Q}}}^X$, of a exciton state, $\vert S,{\boldsymbol{Q}}\rangle$ is obtained by solving the DFT-based BSE.
Fig.  2 presents the calculated low-lying band dispersions, $E_{S,{\boldsymbol{Q}}}^X$, of exciton, including the longitudinal (L) upper bands and transverse (T) lower bands of BX (denoted by B$_u$ and B$_l$, respectively), and those of spin-forbidden GX and DX, of un-strained and strained hBN-encapsulated \ch{MoS2}-MLs with $\varepsilon_{xx}=+5\%$, $\varepsilon_{xx}=0\%$, $\varepsilon_{xx}=-5\%$, respectively,  versus ${\boldsymbol{Q}}$ parallel {\color{black} (Fig.  2 (a),(c) and (e))} or perpendicular {\color{black} (Fig.  2 (b),(d) and (f))} to the uni-axial stress axis. 
{\color{black} For reference, Fig. S1 in the SM \cite{SM} presents the calculated exciton band dispersions of a free-standing \ch{MoS2}-ML, which are qualitatively similar to the exciton band structures of Fig.2. but, due to weakened screening, featured with the fine structure splitting by about two times of magnitude larger than those of hBN-encapsulated \ch{MoS2}-MLs.}

The calculated exciton fine structure of un-strained \ch{MoS2}-ML as shown in Fig.  2 (c) and (d) exhibits the BX-DX splitting of 14 meV, {\color{black}consistent with the experimental observation of Ref.~{\color{black}\citenum{B-D-splitting}}.  This splitting, caused by the combined effects of spin-orbit interaction and {\color{black}Coulomb interactions} \cite{WH-2023,JD-2023,Louie-2015}, is dominated by short-ranged EHEI (SR-EHEI), raising the spin-allowed BX doublet and leaving the spin-forbidden exciton GX and DX as the lowest exciton states.} {\color{black}The SR-EHEI leads to a tiny sub-nm-scaled splitting of the spin-forbidden exciton doublet, and, combined with the spin-orbit interaction, activates the brightness of upper spin-forbidden states with small out-of-plane dipoles. \cite{X-Marie-2016} Hence, the upper band of the DX doublet is often referred to as GX band. {\color{black}Fig.  3 (c)} shows the out-of-plane dipoles of the finite-momentum GX states of an un-strained \ch{MoS2}-ML. Note that, by contrast to the optical activation of the upper GX band, the lower DX band still remain totally dark in a strain-free \ch{MoS2}-ML. }

{\color{black}For un-strained \ch{MoS2}}, it is known that the BX bands are doubly valley-degenerate at the $\Gamma_{ex}$ point where ${\boldsymbol{Q}}=(0,0)$ and, with increasing $\left|\boldsymbol{Q}\right| {\color{black}=Q}\neq 0 $, turn out to be split by the long-ranged {\color{black}EHEI (LR-EHEI)\cite{Louie-2015,Yao-Wang-2014}}. {\color{black}  Note that the upper ${\text B}_u$-band exhibits an unusual quasi-linear dispersion as a consequence of the dominant EHEI of BX that are linearly $Q$-dependent due to the in-plane dipole nature.}\cite{Yao-Wang-2014,GH-2019,Louie-2015}
As the result of LR-EHEI, the BX states in the linear B$_u$-band (parabolic B$_l$-band) carry the longitudinal (transverse) in-plane dipoles whose orientation are aligned (perpendicular) to the exciton momentum. Fig.  3 presents the momentum-dependent transition dipoles of excitons, $\boldsymbol{D}^X_{S,\boldsymbol{Q}}$, in a un-strained and strained \ch{MoS2}-MLs, which are calculated by using the DFT-based exciton theory.\cite{MME-2001,GH-2019,Oscar-2024}


{\color{black}Fig.  2 (a) and (b)} show the calculated exciton band dispersions of a \ch{MoS2}-ML under a fixed tensile uni-axial strain of $\varepsilon_{xx}= + 5\%$ with respect to $\boldsymbol{Q}$ along the $x$- and $y$-axes, respectively. 
First, one notes that, under the tensile uni-axial strain, the parabolic transverse (blue curve) band and the linear longitudinal (red curve) one of BX no longer remain degenerate at the $\Gamma_{ex}$ point, as previously pointed out in the literature. \cite{main-PRB-2022-splitting,Malic-2022,experiment-splitting} 
And, unlike the un-strained case, {\color{black}Fig.  2 (a) and (b)} exhibit distinct exciton band dispersions, indicating the strain-induced anisotropy of the exciton band dispersion.

Note that, in a non-interacting scheme, the strain-induced unequal conduction and valence VDs as shown in Fig. 1 {\color{black}(g)} tend to lower the energy of finite-momentum exciton with increasing the $Q (\le Q_c < \Delta k_{cv})$ in the LC. 
{\color{black} Disregarding the EHEI, the inset of Fig. 2 (a) shows the calculated exciton band structure of the tensilly strained \ch{MoS2}-ML, which indeed exhibits the strain-induced momentum shift of the exchange-free parabolic bands of the both spin-allowed (gray curves) and -forbidden (black curves) exciton. However, the EHEI in exciton that is essentially a dipole-dipole interaction and associated with the strain-modulated exciton dipole acts as another key mechanism to, under the competitive interplay with the strain-induced kinetic VD, affect the strain-modulated exciton band dispersions as well. }

{\color{black} Fig. 2 (a) shows that, despite strain-induced VD }, the lowest BX states (red and blue curves) of a strained \ch{MoS2}-ML stay at the $\Gamma_{ex}$ point, without momentum drift actually. This is because tightly bound BXs in a 2D material possess large in-plane dipoles and are subjected to the strong momentum-dependent EHEI that overwhelms the kinetic VD to dictate the exciton dispersions. By contrast, {\color{black} the strain-induced kinetic VD dominates the dispersive features of the GX and DX bands, the former (latter) of which possesses very small (vanishing) dipoles. 

Fig. 2 (a) shows that the exciton band dispersions of GX and DX preserves the VD-characteristics, similar to the exchange-free structures shown by the inset, but are featured with an anti-crossing, gapped by $\sim 0.1$meV, between the GX and DX bands at $Q=0$ that is caused by the strain-induced short-ranged EHEI. The strain-induced anti-crossing as a result of the combined effect of strain-induced VD and EHEI leads to the unusually negative effective mass of the lowest DX states at $Q\sim 0$ around the $\Gamma_{ex}$ point. }
{\color{black}  As a result}, the spin-forbidden DX band of a MoS2-ML with uni-axial tensile strain $\epsilon_{xx}= 5\%$ shown by Fig.~2 (a) is \st{drastically} reshaped from a parabola to a Mexican-hat-like dispersion 
featured with the unusual sign-reversed effective mass of exciton ($M_{xx}^{\rm{DX}} = 1.265m_0 \rightarrow -0.150m_0$) and the finite-momentum exciton ground states with $Q\sim 20Q_c$.
On the other hand, the GX band of the strained \ch{MoS2}-ML with ${\color{black}\varepsilon_{xx}}= 5\%$ yet remains parabolic and turns out to be featured by drastically reduced effective mass by uni-axial strain ($M_{xx}^{\rm{GX}} = 1.455m_0 \rightarrow 0.125 m_0$), where $M^{\text{S}}_{ij}\equiv \left( \frac{1}{\hbar^2}\frac{\partial^2}{\partial{Q_i}\partial{Q_j}} E_{\text{S},\boldsymbol{Q}}^{X}\right)^{-1}\bigg|_{\boldsymbol{Q} = \boldsymbol{0}}$, and $m_0=9.1\times 10^{-31}$kg is the mass of free electron. More detailed discussion on strain-modulated exciton masses will be given later by Sec. D.


Then, let us proceed to examine the the exciton band dispersion of a \ch{MoS2}-ML under compressive uni-axial strain, as shown in {\color{black}Fig.  2 (e) and (f)}.
 Comparing {\color{black}Fig.  2 (e) and (f)} with {\color{black}Fig.  2 (a) and (b)}, one observes that the exciton band dispersion of a \ch{MoS2}-ML with tensile and compressive strains follow quite similar dispersions, except for the reversed order of the transverse and longitudinal BX bands. 
Such a similarity is straightforward to understand, since, due to the non-zero Poisson's ratio, applying a imposed compressive uni-axial strain to a 2D material always accompanies with another tensile uni-axial strain perpendicular to the uni-axial strain axis, and vice versa.
{\color{black} Re-orientating the uni-axial strain to the armchair direction ($y$-direction), one can show  (See Fig.S2 in SM \cite{SM}) that the strain-modulated exciton band structures of a strained \ch{MoS2}-ML remain similar, both qualitatively and quantitatively, to those of a MoS2 monolayer under an uni-axial strain along the x-axis. Interestingly, under an uni-axial strain in the y-direction, the momentum shift of the exciton band edges yet remain towards the x-direction, same as the cases of uni-axial strain along the x-direction. This is because, as pointed out by {\color{black} Ref.~\citenum{valley-drift-2019}}, the strain-induced VD is always along the zigzag-direction ($x$-direction) no matter the uni-axial strain is along the zigzag or armchair directions.  
Thus, hereafter we shall focus the investigation only on the \ch{MoS2}-MLs with tensile uni-axial strains along the zigzag axis ($x$-direction) to reduce the repetition of physical analysis.}

 \subsection{Strain-modulated exciton dipoles and band energy contours}
Fig.  3 (a-d) [(e-h)] present the energy contour plots of the calculated exciton bands of the \ch{MoS2}-MLs without strain [with tensile strain]. The transition dipole, $\boldsymbol{D}^X_{S\boldsymbol{Q}}$, of an exciton state with the momentum $\hbar \boldsymbol{Q}$ is represented by a line segment (representing a linearly polarized dipole) or a circle (representing a circularly polarized out-of-plane dipole) placed at the position of $\boldsymbol{Q}$ in the momentum plane, whose size and orientation reflect the magnitude and orientation of dipole, respectively. Since the lowest DX band of an un-strained \ch{MoS2}-ML are optically completely dark, no symbol of dipole appear in {\color{black}Fig.  3 (d)}.

In Fig. 3 (a-d), all exciton bands of an un-strained \ch{MoS2}-ML remain nearly isotropic, and each finite momentum exciton states hold well-defined L, T, or out-of-plane dipoles. 
{\color{black} By contrast, Fig. 3 (e-h) shows the diversified anisotropies for the exciton band dispersions of an uni-axially strained \ch{MoS2}-ML. In Fig. 3 (e) and (f), the transition dipoles of the BX states are shown to be L-T-mixed.} Futhermore, our {\color{black}DFT} studies reveal that the application of a uni-axial stress onto a TMD-ML impacts actually most the spin-forbidden exciton states, including the band dispersions and the optical selection rules as well.
{\color{black}In Fig.  3 (h)}, we see that the spin-forbidden DX states with finite momenta around the direction along the stress-axis exhibit small strain-induced out-of-plane dipoles, leading to weak brightness and indicating the violation of the spin selection rule for DX. The strain-induced optical brightness in the DX states results from the symmetry breaking caused by the uni-axial strain, {\color{black} which, as in the pseudo-spin model for exciton doublet presented by Ref.~\citenum{1-D-G-mixing}, gives rise to the effective pseudo-magnetic field component that couples the DX to the GX states with small out-of-plane dipoles.} \cite{1-D-G-mixing} 
{\color{black} Because of the out-of-plane dipole nature, the EHEI of the GX states with $\boldsymbol{Q}=\boldsymbol{0}$, {\color{black} unlike the case of BX}, is non-vanishing and remain roughly constant with varying $\boldsymbol{Q}$.
As detailed later, the combined effect of strain-induced VD and the non-vanishing EHEI at $\boldsymbol{Q}=\boldsymbol{0}$ retains the parabolicity in the band dispersion of GX featured with strain-reduced effective mass (See {\color{black}Fig.~4 (a)}), and tends to maintain the isotropy of the optical pattern of GX band against the imposed uni-axial strain (See {\color{black}Fig.~5 (i)}). }


\subsection{Strain-diversified exciton masses}
As a result of strain-induced VD and non-vanishing EHEI at $\boldsymbol{Q}=\boldsymbol{0}$, the curvature of the upper parabolic GX band along the $Q_x$-axis is significantly increased and the effective mass of the GX band along the $x$-axis is dropped from  $M_{xx}^{\text{GX}}= 1.455 m_0 $ to  $0.125 m_0 $ (See Fig.  2 (c)), by one order of magnitude, with varying $\varepsilon_{xx}=0\% \rightarrow +5\%$. Following the studies of Refs.~\citenum{transport-1993,1-exciton-transport,2-exciton-transport,3-exciton-transport}, the directional steady diffusivity of exciton is directly quantified by the {\color{black}diffusion coefficient $\boldsymbol{D}^{\rm diff}$ whose elements are proportional to the inverse exciton effective mass, {\it i.e.},
$D_{ij}^{\rm diff}  \propto \left( M_{ij}^S \right)^{-1}$, $i,j=x\text{ or } y$.}

\begin{figure}
\centering 
\includegraphics[width=1\textwidth]{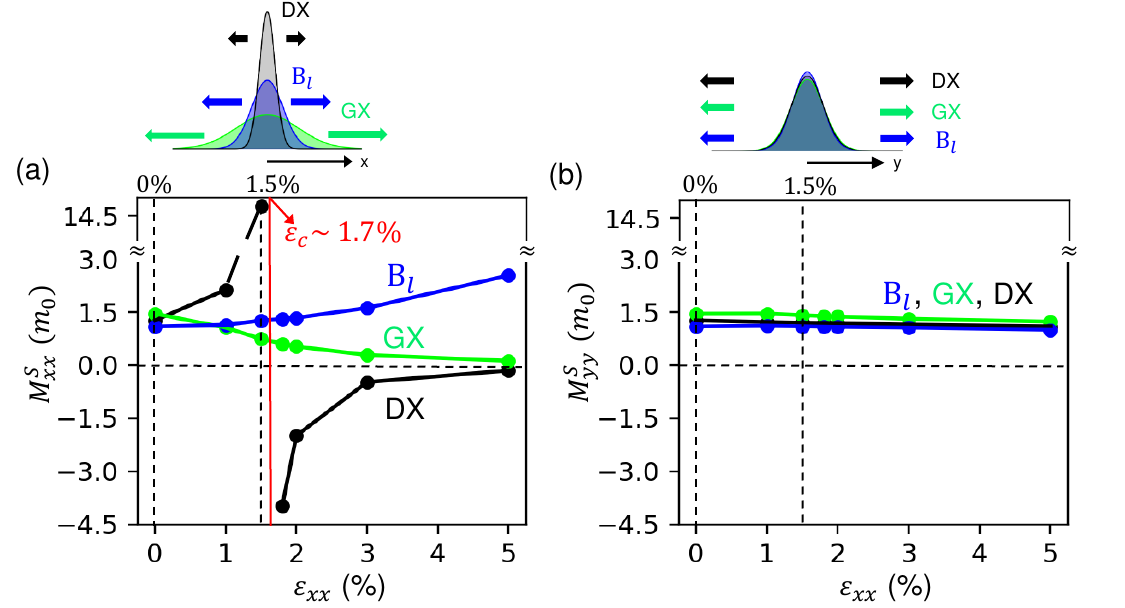}
\caption{(a) [(b)] The strain-modulated effective masses, $M_{xx}^S$ [$M_{yy}^S$], of BX (blue curve), GX (green curve), and DX (black curve) along the $x$-axis [$y$-axis] of a MoS$_2$-ML under the uni-axial strain varied from $\varepsilon_{xx}=0\%$ to $\varepsilon_{xx}=+5\%$. Inset:  Schematics of exciton diffusion of BX, GX and DX in an uni-axially strained MoS$_2$-ML with $\epsilon_{xx}=1.5\%$, along the $x$-direction [$y$-direction] with diversified [similar] diffusive lengths.}
\label{Fig.  4a} 
\label{Fig.  4c} 
\label{Fig.  4b}
\end{figure}


Fig.  4 (a) presents the strain-dependences of the effective masses of excitons along the uni-axial strain axis, $M^{S}_{xx}$, for longitudinal BX, GX, and DX ($S=\text{B}_l,$ GX, and DX) in a strained \ch{MoS2}-ML with varying the tensile uni-axial strain, which are shown very diversified. With increasing the uni-axial strain from $\varepsilon_{xx} = 0\%$ up to $\varepsilon_{xx} = +5\%$, one sees that the $M^{\text{B}_{l}}_{xx}$ gradually increases from $1.094m_0$ to $2.535m_0$ but, oppositely, the $M^{\text{GX}}_{xx}$  decreases drastically from $1.455m_0$ down to $ 0.125m_0$. The presence of uni-axial strain is shown to impact most the $M^{\text{DX}}_{xx}$ as presented by the black curve in {\color{black}Fig.  4 (a)}. With increasing $\varepsilon_{xx}$, $M^{\text{DX}}_{xx}$ quickly increases but abruptly reverses its sign to become negative at the critical strain {\color{black}$\varepsilon_{xx}\approx1.7\%\equiv \varepsilon_{c}$}, where the Mexican-hat-like band dispersion emerges. With the strain greater than $\varepsilon_{c}$, the negative $M^{\text{DX}}_{xx}$ yet turns out to decrease its magnitude. The non-monotonic strain-dependence of $M^{\text{DX}}_{xx}$ results from the emergence of the unusual strain-induced Mexican-hat-like dispersion, manifesting the pronounced excitonic effect of VD on the spin-forbidden exciton.

By contrast to strain-diversified exciton masses along the $x$-direction, $M^{S}_{xx}$, the effective masses of exciton along the $y$-axis, $M^{S}_{yy}$, remain almost unchanged against varied $\varepsilon_{xx}$, as one sees in Fig.  4 (b). This indicates the strain-insensitive exciton diffusion along the direction perpendicular to the axis of uni-axial strain, and the strain-induced anisotropy in the exciton diffusivity. Along the uni-axial strain axis, the strain-diversified exciton masses, $M^{S}_{xx}$, of an uni-axially strained TMD-ML should lead to distinct diffusion lengths of BX, GX, and DX, especially as  $\epsilon_{xx} \sim \epsilon_c$, as schematically shown by Fig.  4.
The predicted diversification in the effective masses of exciton reveals the possibility to spatially resolve the BX, GX, and DX in exciton transport experiments by imposing an uni-axial strain to a TMD-ML.

\begin{figure}
\centering 
\includegraphics[width=\textwidth]{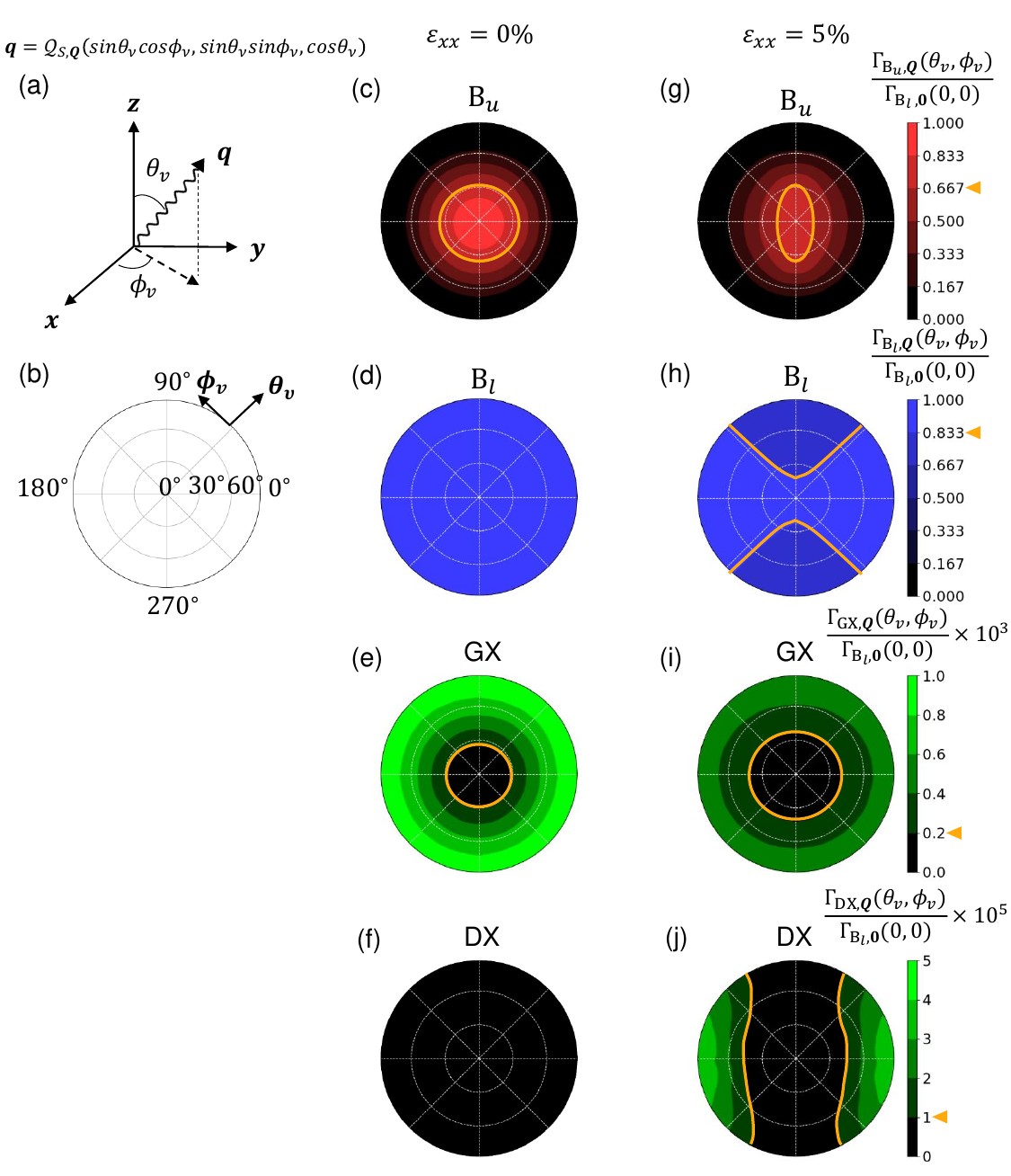} 
\caption{ (a) Schematics of a directional light beam of PL with the polar angle, $\theta_{\nu}$, and the azimuthal angle, $\phi_{\nu}$, in the spherical coordinate, and $\mathcal{Q}_{S,\boldsymbol{Q}}=\frac{E^X_{S,\boldsymbol{Q}}}{\hbar c}$ represents  the magnitude of the wave vector of the emitted light. (b) The top view for a hemisphere, where the radius of circle represents $\theta_\nu$ and the azimuthal angle represents $\phi_\nu$. (c-f) The $\theta_{\nu}$- and $\phi_{\nu}$-resolved optical patterns of the relative transition rates, $\Gamma_{S ,\bf{Q}}^S(\theta_{v},\phi_{v})/ \Gamma_{B_{l}, \bf{0}}^S(0,0)$, of the directional PLs from the B$_u$, B$_l$, GX, and DX exciton states, respectively, of an un-strained \ch{MoS2}-ML. (g-j) The angle-resolved optical patterns of the four types of exciton states of an uni-axially strained \ch{MoS2}-ML with $\varepsilon_{xx}=5\%$.
{\color{black}The orange-coloured energy contours illustrates the strain-induced anisotropies of the directional optical patterns of \ch{MoS2}-ML.}
}
\label{Fig.  5} 
\end{figure}

\subsection{Angle-resolved photoluminescence spectra}

The strain-induced diversification and anisotropies in the exciton band dispersion of BX, GX, and DX  leads also to the diversified angle-dependences of the directional photo-luminescences (PLs) from the different types of exciton in an uni-axially strained \ch{MoS2}-ML. 
Based on the solved exciton states and momentum-dependent transition dipoles of BX, GX, and DX, we calculate the transition rates of the directional PLs from an exciton in an uni-axially strained \ch{MoS2}-ML, as a function of the polar ($\theta_{\nu}$) and azimuthal ($\phi_{\nu}$) angles of the emitted light, {\color{black} using the formalisms of {\color{black}Appendix C} based on the Fermi's golden rule}. 

Fig.  5 shows the optical patterns of the directional PLs, characterized by the momentum-dependent transition rates of exciton, $\Gamma_{\rm{S},\boldsymbol{Q}}(\theta_v, \phi_v) $, from the exciton states of un-strained and uni-axially strained \ch{MoS2}-MLs. 
Fig.  5 (c-f) present the optical patterns of the $\rm{B}_u$, $\rm{B}_l$, GX, and DX bands of an un-strained \ch{MoS2}-ML, all of which present isotropic features. 
In the presence of uni-axial strain, the optical patterns of the BX states of an uni-axially strained \ch{MoS2}-ML turn out to be anisotropic, as shown by Fig.  5 (g) and (h). 
By contrast to the BX states, the optical pattern of the GX states of a uni-axially strained \ch{MoS2}-ML yet is found strain-insensitive and presents only slight anisotropy as one sees in Fig.  5 (i). {\color{black}  This is because the the {\it e-h} exchange interaction of the GX state with out-of-plane dipoles is insensitive to the applied in-plane strain. }
 Interestingly, we show that the optical pattern of the DX states exhibit the most anisotropy (See Fig.  5 (j)).
The strain-induced small brightness of DX yields the weak directional light emission that tends to propagate in the direction around the $x$-axis, nearly horizontally with large polar angle. The large-polar-angle directional light emission from the long-lived spin-forbidden DXs in a uni-axially strained \ch{MoS2}-ML could be considered as a  signature of strain-induced excitonic effect of VD, optically measurable in the time- and angle-resolved optical spectroscopy.  

\section{Conclusions}
In summary, we have presented a comprehensive theoretical investigation of the strain-modulated excitonic fine structures and band dispersions of uni-axially strained \ch{MoS$_2$}-MLs by solving the first-principles-based BSE. As a main finding,  imposing an uni-axial strain onto a \ch{MoS2}-MLs significantly diversifies the band dispersions, diffusive transport properties, and angle-resolved optical patterns of the BX, GX, and DX states, as a consequence of the competitive interplay between the strain-induced VD and momentum-dependent EHEI. 
We show that, while the band dispersions of BX doublet remain almost unchanged against strain, the presence of an uni-axial strain in a \ch{MoS2}-ML impacts most the spin-forbidden exciton states, both GX and DX. 
Imposing only a few-percentage uni-axial strain onto a \ch{MoS2}-ML can reshape the band dispersion of DX states from a parabola to a Mexican-hat-like profile, featured with strain-activated brightness and abrupt sign-reversal of the effective mass of DX. Oppositely, the band dispersion of GX in a uni-axial strained \ch{MoS2}-ML that remain parabolic is featured by a drastically strain-reduced effective mass. The strain-induced diversification in the exciton band dispersions of a uni-axial strained \ch{MoS2}-ML is shown to diversify and make very distinct the diffusive lengths and angle-resolved optical patterns of BX, GX, and DX as well, as the measureable signatures of strain-induced VD.
This suggests the possibility of not only {\it spectrally} but also {\it spatially} resolving the BX, GX and DX exciton states of a uni-axially strained 2D material in exciton diffusive transport experiments or the angle-resolved optical spectroscopy.



\section*{ACKNOWLEDGMENTS}
\noindent

The research study is supported financially by the NSTC of Taiwan (Grant Nos. NSTC 112-2112-M-A49-019-MY3, NSTC 112-2622-M-A49-001, NSTC 113-2112-M-A49-035-MY3, and NSTC 113-2124-M-A49-008), and  T-Star center project “Future Semiconductor Technology Research Center”, under Grant No. NSTC 113-2634-F-A49-008. The authors thank the National Center for High-Performance Computing (NCHC) in Taiwan for providing computational resources.

\section*{DATA AVAILABILITY}
The data that support the findings of this article are not publicly available. The data are available from the authors upon reasonable request.

\appendix
{\color{black}
\section{Technical information of the DFT calculations}
 In the DFT calculation, we take 15 $\times$ 15 $\times$  1 Monkhorst-Pack k-grid that ensures the convergence of our calculation.  For the simulation of a 2D \ch{MoS2}-ML, we set up the \ch{MoS2}-ML super-cells with the sufficiently thick vacuum layer between periodic \ch{MoS2}-MLs, 31.8 ${\color{black}\mathring{A}}$, to avoid the inter-layer coupling. We adopt the Perdew-Burke-Ernzerhof (PBE)\cite{PBE} model for the exchange-correlation functional and the norm-conserving pseudopotential which is generated with the code ONCVPSP (Optimized Norm-Conserving Vanderbilt Pseudopotential). The plane-wave cutoff energy is set to 70 Ry.
For a un-strained \ch{MoS2}-ML, we adopt the lattice constant $a_{0}=3.16 {\color{black}\mathring{A}}$ in the DFT calculation.

}
\section{Exciton dipole moment}
The transition dipole moment of an exciton in the state $\vert S,\boldsymbol{Q}\rangle$  reads
\begin{equation}
 \boldsymbol{D}_{S,\boldsymbol{Q}} =\frac{1}{\sqrt{\mathcal{A}}}\sum_{vc\boldsymbol{k}}A_{S,\boldsymbol{Q}}(vc\boldsymbol{k})\boldsymbol{d}_{v,\boldsymbol{k}; c,\boldsymbol{k}},
\end{equation}
where $\boldsymbol{d}_{v,\boldsymbol{k}; c,\boldsymbol{k}}=e\langle{\psi_{v,\boldsymbol{k}} }|\boldsymbol{r}|{\psi_{c,\boldsymbol{k}} }\rangle$ is the dipole moment of single-particle transition. \cite{GH-2019,Oscar-2024,PY-2021}
The single-particle dipole moment $\boldsymbol{d}_{v,\boldsymbol{k}; c,\boldsymbol{k}}$ can be related to the momentum matrix element $\boldsymbol{p}_{v,\boldsymbol{k}; c,\boldsymbol{k}}$ \cite{MME-2001} through the commutator relation, $\boldsymbol{p} = \frac{i m_0}{\hbar} [H^{\rm{KS}}, \boldsymbol{r}]$  \cite{GH-2019,Oscar-2024,PY-2021}, and reformulated as

\begin{equation}
\boldsymbol{d}_{v,\boldsymbol{k}; c,\boldsymbol{k}}=e\langle{\psi_{v,\boldsymbol{k}} }|\boldsymbol{r}|{\psi_{c,\boldsymbol{k}} }\rangle=\frac{e}{\epsilon_{v,\boldsymbol{k}}-\epsilon_{c,\boldsymbol{k}}}\left[\frac{\boldsymbol{p}_{v,\boldsymbol{k}; c,\boldsymbol{k}^{\prime}}}{\frac{im_{0}}{\hbar}}\right].
\end{equation}
Thus, the momentum matrix element can be explicitly expressed in terms of MLWFs and WTB-Hamiltonian matrix element as
\begin{equation}
\begin{split}
\boldsymbol{p}_{v,\boldsymbol{k}; c,\boldsymbol{k}}&=\frac{im_{0}}{\hbar}\left(\epsilon_{v,\boldsymbol{k}}-\epsilon_{c,\boldsymbol{k}}\right)\sum_{j,j^{\prime}}\sum_{\boldsymbol{R}} C_{j^{\prime}}^{\left(v\right)\ast} \left(\boldsymbol{k}\right) C_j^{\left(c\right)} \left(\boldsymbol{k}\right)e^{i\boldsymbol{k}\cdot \boldsymbol{R}}
\langle{\mathcal{W}_{j^{\prime},\boldsymbol{0}} }|\boldsymbol{r}|{\mathcal{W}_{j,\boldsymbol{R}} }\rangle
\\
&+\frac{m_{0}}{\hbar}\sum_{j,j^{\prime}} C_{j^{\prime}}^{\left(v\right)\ast} \left(\boldsymbol{k}\right) C_j^{\left(c\right)} \left(\boldsymbol{k}\right) \nabla_{\boldsymbol{k}}
H_{j^{\prime} j}^{TB}(\boldsymbol{k})
\end{split}
\end{equation}
where the first term corresponds to intra-atomic transitions $ \boldsymbol{p}^{intra}_{v,\boldsymbol{k}; c,\boldsymbol{k}}$, while the second term corresponds to inter-atomic transitions $\boldsymbol{p}^{inter}_{v,\boldsymbol{k}; c,\boldsymbol{k}}$. Both terms can be evaluated from the WTB-Hamiltonian matrix $H^{\rm{TB}}(\boldsymbol{k})$ and the MLWFs using the WANNIER90 package.

\section{ Angle-resolved photoluminescence spectra}

\subsection{Hamiltonian of light-matter interaction}
The Hamiltonian of light-matter interaction between an electron and a EM wave is given by
\begin{equation}
\tilde{H}'(\boldsymbol{r}, t) = \frac{|e|}{m_0} \boldsymbol{A}(\boldsymbol{r}, t) \cdot \boldsymbol{p},
\end{equation}
where $\boldsymbol{A}(\boldsymbol{r}, t)$ is the space- and time-varying vector potential of the EM wave.
In quantum electrodynamics, one expresses the time-dependent vector potential in the language of second quantization as

\begin{equation}
\hat{A}(\boldsymbol{r}, t) = \sum_{\boldsymbol{q}} \sum_{\lambda} \sqrt{\frac{\hbar}{2 \epsilon_0 \omega_{\boldsymbol{q}} V}} \left[ \hat{a}_{\boldsymbol{q},\lambda}  e^{i(\boldsymbol{q} \cdot \boldsymbol{r} - \omega_{\boldsymbol{q}} t)} + \hat{a}_{\boldsymbol{q},\lambda}^{\dagger}  e^{-i(\boldsymbol{q} \cdot \boldsymbol{r} - \omega_{\boldsymbol{q}} t)} \right] \hat{\boldsymbol{e}}_{\boldsymbol{q}, \lambda}.
\end{equation}
where $\boldsymbol{q}= \mathcal{Q}\left( \sin\theta_{\nu} \cos\phi_{\nu} \hat{\boldsymbol{x}} + \sin \theta_{\nu} \sin\phi_{\nu} \hat{\boldsymbol{y}} + \cos\theta_{\nu}\hat{\boldsymbol{z}} \right)$ denotes the wave vector of a directional light beam in the polar coordinate,  $\mathcal{Q}$ is the magnitude of the wave vector of light, $\hat{\boldsymbol{e}}_{\boldsymbol{q}, \lambda=1,2}$ is the transverse-polarization basis of the light that is required to satisfy $\hat{\boldsymbol{e}}_{\boldsymbol{q}, \lambda} \perp \boldsymbol{q}$. By convention, we adopt $\hat{\boldsymbol{e}}_{\boldsymbol{q}, \lambda=1}= \cos\theta_{\nu} \cos\phi_{\nu} \hat{\boldsymbol{x}} + \cos \theta_{\nu} \sin\phi_{\nu} \hat{\boldsymbol{y}} - \sin\theta_{\nu}\hat{\boldsymbol{z}}$, and $\hat{\boldsymbol{e}}_{\boldsymbol{q}, \lambda=2}= -\sin\phi_{\nu} \hat{\boldsymbol{x}} + \cos\phi_{\nu} \hat{\boldsymbol{y}}$, ensuring $\boldsymbol{q} \perp \hat{\boldsymbol{e}}_{\boldsymbol{q}, \lambda=1} \And \boldsymbol{q} \perp \hat{\boldsymbol{e}}_{\boldsymbol{q}, \lambda=2} \And \hat{\boldsymbol{e}}_{\boldsymbol{q}, \lambda=1} \perp \hat{\boldsymbol{e}}_{\boldsymbol{q}, \lambda=2}$ (as illustrated by Fig. 6), $V$ is the volume of the material, $\epsilon_0$ is the vacumn permittivity, and $\omega_{\boldsymbol{q}}$ is the angular frequency of light.

For a light beam emitted from an exciton in the state $\vert S,\boldsymbol{Q} \rangle$, the magnitude of the wave vector of the light is $ \mathcal{Q}_{S,\boldsymbol{Q}}=\frac{ E_{S\boldsymbol{Q}}^X}{\hbar c}$ that satisfies the energy conservation law. 
Accordingly, the time-dependent Hamiltonian in second quantization is expressed as
\begin{equation}
{\color{black}\hat{\tilde{H'}}_ {\hat{\boldsymbol{e}}_{\boldsymbol{q}, \lambda},{\boldsymbol{q}}} (\boldsymbol{r}, t) =  \sum_{\boldsymbol{q}}\sum_{\lambda}  \left[ \tilde{M}_{i, j}^{\hat{\boldsymbol{e}}_{\boldsymbol{q}, \lambda},{\boldsymbol{q}}} \hat{a}_{\boldsymbol{q},\lambda} \hat{c}_{c,\boldsymbol{k}+\boldsymbol{q}}^\dagger \hat{c}_{v,\boldsymbol{q}}e^{-i\omega_{\boldsymbol{q}} t} + \tilde{M}_{i, j}^{\hat{\boldsymbol{e}}_{\boldsymbol{q}, \lambda},{\boldsymbol{q}}} \hat{a}_{\boldsymbol{q},\lambda}^\dagger  \hat{c}_{c,\boldsymbol{k}+\boldsymbol{q}}  \hat{c}_{v,\boldsymbol{q}}^\dagger e^{i\omega_{\boldsymbol{q}} t} \right]}  \, ,
\end{equation}
where the optical matrix elements
$\tilde{M}_{i,j}^{\hat{\boldsymbol{e}}_{\boldsymbol{q}, \lambda}, \boldsymbol{q}} \equiv \langle \psi_i \vert \tilde{H'}_{\hat{\boldsymbol{e}}_{\boldsymbol{q}, \lambda}, \boldsymbol{q}} (\boldsymbol{r}) \vert \psi_j \rangle =  {\color{black} \frac{\vert e\vert }{ m_0}\sqrt{\frac{\hbar}{2\epsilon_0 \omega_{\boldsymbol{q}} V}} \hat{\boldsymbol{e}}_{\boldsymbol{q}, \lambda}\cdot \langle \psi_i \vert \boldsymbol{p} e^{i\boldsymbol{q}\cdot \boldsymbol{r}} \vert \psi_j \rangle} $. 
\par
\begin{figure} 
\includegraphics[width=\textwidth]{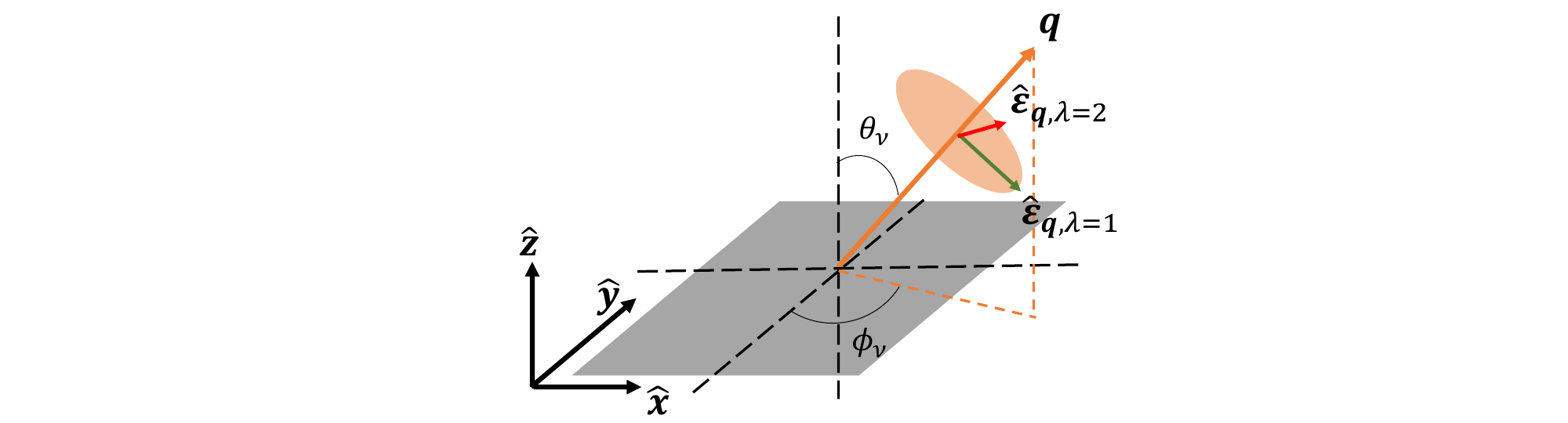}

\caption{The schematic illustrates the polar angle, $\theta_{\nu}$,  azimuthal angle, $\phi_{\nu}$, and the transverse polarization basis, , $\hat{\boldsymbol{e}}_{\boldsymbol{q}, \lambda=1,2}$, of a directional light beam with the wave vector $\boldsymbol{q}$ emitted from a \ch{MoS2}-ML.}
\end{figure}
\subsection{ Angle-dependent transition rate of directional PL from an exciton}
Following the Fermi's golden rule, the transition rate of spontaneous emssion (SE) from an exciton state $|S,\boldsymbol{Q}\rangle$ to $|\rm{GS}\rangle$ is evaluated by  
\begin{equation}
\Gamma^{\hat{\boldsymbol{e}}_{\boldsymbol{q}, \lambda},\boldsymbol{q}}_{S,\boldsymbol{Q}} =\frac{2\pi}{\hbar} \lvert \tilde{M}^{\hat{\boldsymbol{e}}_{\boldsymbol{q}, \lambda},\boldsymbol{q}}_{S,\boldsymbol{Q}}\rvert^2 \delta(E^{X}_{S,\boldsymbol{Q}}-\hbar\omega_{\boldsymbol{q}}),
\end{equation}
where 
\begin{equation}
\tilde{M}_{S,\boldsymbol{Q}}^{\hat{\boldsymbol{e}}_{\boldsymbol{q}, \lambda}, \boldsymbol{q}} = \delta_{\boldsymbol{q}^{\parallel}, \boldsymbol{Q} } \ {\color{black} \frac{E_g}{i\hbar}  \sqrt{\frac{\hbar}{2\epsilon_0 \omega_{\boldsymbol{q}} V}} \left(  \hat{\boldsymbol{e}}_{\boldsymbol{q},\lambda }\cdot \boldsymbol{D}_{S,\boldsymbol{Q}}^{X} \right)\,},
\label{MME}
\end{equation}
where the momentum conservation law ensures that $\boldsymbol{q}^{\parallel}= \mathcal{Q}_{S,\boldsymbol{Q}}\left( \sin\theta_{\nu} \cos\phi_{\nu} \hat{\boldsymbol{x}} + \sin \theta_{\nu} \sin\phi_{\nu} \hat{\boldsymbol{y}}  \right) = \boldsymbol{Q}$. 
Summing over all the modes of polarization $\lambda$ and wave vectors $\boldsymbol{q}$, we can obtain the total SE rate of the directional PLs from an exciton state as follows
\begin{equation}
\Gamma_{S,\boldsymbol{Q}}\left(\theta_\nu,\phi_\nu \right) =\frac{V}{(2\pi)^3}\int^\infty_0 dq \ q^2 \sum_{\lambda}\Gamma^{\hat{\boldsymbol{e}}_{\boldsymbol{q}, \lambda},\boldsymbol{q}}_{S,\boldsymbol{Q}}=\frac{V}{\left(2\pi\right)^2}\frac{\mathcal{Q}^2_{S,\boldsymbol{Q}}}{\hbar^2c}\sum_{\lambda} \lvert \tilde{M}_{S,\boldsymbol{Q}}^{\hat{\boldsymbol{e}}_{\boldsymbol{q}, \lambda} \boldsymbol{q}} \rvert^2,
\label{Gamma_SQ}
\end{equation}
In Eq.~\ref{Gamma_SQ}, the angle-dependences of $\Gamma_{S,\boldsymbol{Q}}\left(\theta_\nu,\phi_\nu \right)$ mainly follow those of the optical matrix element, $\tilde{M}_{S,\boldsymbol{Q}}^{\hat{\boldsymbol{e}}_{\boldsymbol{q}, \lambda},\boldsymbol{q}}$ 
, which is governed by the exciton dipole, $ \boldsymbol{D}_{S,\boldsymbol{Q}}^{X}$, whose magnitude and orientation depend on the type of exciton.
By substituting Eq.~\ref{MME} into Eq.~\ref{Gamma_SQ}, one can find the the angle-dependence of $\Gamma_{S,\boldsymbol{Q}}\left(\theta_\nu,\phi_\nu \right)$ is dominated by $ \boldsymbol{D}_{S,\boldsymbol{Q}}^{X}$. The calculated $\Gamma_{S,\boldsymbol{Q}}\left(\theta_\nu,\phi_\nu \right)$ of a uni-axially strained \ch{MoS2}-ML are shown in Fig. 5 (g-j) .

\end{document}